\documentclass[graybox]{svmult}

\usepackage{natbib}
\usepackage{mathptmx}       
\usepackage{helvet}         
\usepackage{hyperref}
\usepackage{courier}        
\usepackage{type1cm}        
\usepackage{makeidx}         
\usepackage{graphicx}        
\usepackage{multicol}        
\usepackage[bottom]{footmisc}
\newcommand{\diamonds}{\textsc{D\large{iamonds}}}
\usepackage{bm}

\makeindex             

\begin{document}

\title*{Asteroseismic Data Analysis with DIAMONDS}
\author{Enrico Corsaro}
\institute{Enrico Corsaro \at INAF - Osservatorio Astrofisico di Catania, via S. Sofia 78, I-95123, Catania, Italy, \\ \email{enrico.corsaro@oact.inaf.it}\\ \\
Laboratoire AIM Paris-Saclay, CEA/DRF - CNRS - Universit\'e Paris Diderot, IRFU/SAp, L'Orme des Merisiers, bat. 709, 91191 Gif-sur-Yvette Cedex, France\\ \\
Instituto de Astrof\'{i}sica de Canarias, E-38205 -- Universidad de La Laguna, Departamento de Astrof\'{i}sica, E-38206, La Laguna, Tenerife, Spain
}
%
%
\maketitle

\abstract{Since the advent of the space-based photometric missions such as \textit{CoRoT} and NASA's \textit{Kepler}, asteroseismology has acquired a central role in our understanding about stellar physics. The \textit{Kepler} spacecraft, especially, is still releasing excellent photometric observations that contain a large amount of information not yet investigated. For exploiting the full potential of these data, sophisticated and robust analysis tools are now essential, so that further constraining of stellar structure and evolutionary models can be obtained. In addition, extracting detailed asteroseismic properties for many stars can yield new insights on their correlations to fundamental stellar properties and dynamics. After a brief introduction to the Bayesian notion of probability, I describe the code \diamonds\,\,for Bayesian parameter estimation and model comparison by means of the nested sampling Monte Carlo (NSMC) algorithm. NSMC constitutes an efficient and powerful method, in replacement to standard Markov chain Monte Carlo, very suitable for high-dimensional and multimodal problems that are typical of detailed asteroseismic analyses, such as the fitting and mode identification of individual oscillation modes in stars (known as peak-bagging). \diamonds\,\,is able to provide robust results for statistical inferences involving tens of individual oscillation modes, while at the same time preserving a considerable computational efficiency for identifying the solution. In the tutorial, I will present the fitting of the stellar background signal and the peak-bagging analysis of the oscillation modes in a red-giant star, providing an example to use Bayesian evidence for assessing the peak significance of the fitted oscillation peaks.}

\section{Bayesian statistics}
\label{sec:1}
Let us assume to consider a given physical problem, e.g., the fitting of an observational dataset through the use of a predictive model. We term the dataset $D$ and the fitting model $\mathcal{M}_k$, the latter having a number of $k$ free parameters that we represent with the $k$-dimensional parameter vector $\bm{\theta} = (\theta_1, \theta_2, \dots, \theta_k)$. The number of free parameters sets the dimensionality of the problem, to which a $k$-dimensional parameter space $\Omega_{\mathcal{M}_k}$ is associated, representing the space of the solutions. Our aim is to obtain optimal estimates of each free parameter and a corresponding statistical weight of the model $\mathcal{M}_k$ that takes into account both the number of dimensions and the fit quality. This statistical inference can be properly addressed through the means of Bayesian statistics \citep{Jeffreys61,Sivia06,Trotta08,Bolstad13,Corsaro13,Corsaro14}. In particular, the core of the statistical representation is given by Bayes' theorem:
\begin{equation}
p (\bm{\theta} \mid D, \mathcal{M}_k) = \frac{\mathcal{L}(\bm{\theta} \mid D, \mathcal{M}_k) \pi (\bm{\theta} \mid \mathcal{M}_k)}{p (D \mid \mathcal{M}_k)} \, ,
\label{eq:bayes}
\end{equation}
where $\mathcal{L}(\bm{\theta} \mid D, \mathcal{M}_k)$ (hereafter, $\mathcal{L}(\bm{\theta})$ for simplicity) is the likelihood function, which represents the way we sample the data, while $\pi (\bm{\theta} \mid \mathcal{M}_k)$ is the prior probability density function (PDF) that reflects our knowledge about the model parameters. The left-hand side of Eq.~(\ref{eq:bayes}) is the posterior PDF, which has a key role in the parameter estimation problem. Through a marginalization of the posterior PDF, namely an integration over the uninteresting free parameters, we estimate the free parameters of the model. Among the different estimators for each parameter, in Bayesian statistics the median is usually preferred because it represents the most resistant estimator, namely the least sensitive to possible outliers, and because it is invariant for variable change.

The denominator on the right-hand side of Eq.~(\ref{eq:bayes}) is instead a normalization factor, generally known as the Bayesian evidence (or marginal likelihood), which is defined as
\begin{equation}
\mathcal{E} \equiv p (D \mid \mathcal{M}_k) = \int_{\Omega_{\mathcal{M}_k}} \mathcal{L}(\bm{\theta} \mid D, \mathcal{M}_k) \pi (\bm{\theta} \mid \mathcal{M}_k) d \bm{\theta} \, .
\label{eq:evidence}
\end{equation}
The Bayesian evidence is used for as a statistical weight for model comparison because it encompasses the principle of the Occam's razor, meaning that models are favored if they provide a better fit to the data but are penalized if their number of free parameters is larger than that of a competitor model. For our study, the model comparison is performed by computation of the Bayes' factor  $\mathcal{B}_{ij} = \mathcal{E}_i / \mathcal{E}_j$ (see also Sect.~\ref{sec:6}), in which the model corresponding to a larger Bayesian evidence is statistically more likely \citep{Jeffreys61,Trotta08,Corsaro13,Corsaro14}.

\section{Nested sampling Monte Carlo}
\label{sec:2}
Since Eq.~(\ref{eq:evidence}) is a multi-dimensional integral, with increasing number of dimensions its evaluation becomes quickly unsolvable
both analytically and by numerical approximations. For overcoming this problem, a NSMC algorithm was developed \citep{Skilling04}. This algorithm allows for
an efficient evaluation of the Bayesian evidence for any number of dimensions and provides the sampling of the posterior probability distribution
(PPD) for parameter estimation as a straightforward byproduct. Detailed descriptions of the algorithm can be found in \citet{Skilling04,Sivia06,Feroz08,Feroz09,Corsaro14}. 

In short, a prior mass $X$ is defined such that
\begin{equation}
X (\mathcal{L}^*) = \int_{\mathcal{L}(\bm{\theta}) > \mathcal{L}^*} \pi (\bm{\theta} \mid \mathcal{M} ) d \bm{\theta} \, ,
\label{eq:prior_mass}
\end{equation}
with $\mathcal{L}^*$ being some fixed value of the likelihood function. As a consequence, $0 \leq X \leq 1$ because $\pi (\bm{\theta} \mid \mathcal{M})$ is a PDF. Equation (\ref{eq:prior_mass}) is
therefore the fraction of volume under the prior PDF that is contained within the hard constraint $\mathcal{L} (\bm{\theta}) > \mathcal{L}^*$. This means that the higher is the constraining value $\mathcal{L}^*$, the smaller is the prior mass considered. This is equivalent to considering a portion of parameter space delimited by the iso-likelihood contour $\mathcal{L} (\bm{\theta}) = \mathcal{L}^*$,
in which also the maximum value $\mathcal{L}_\mathrm{max}$ is contained.

In the NSMC, the sampling of the posterior PDF is performed by starting with a prior mass $X = 0$ (thus considering the entire parameter space) and an initial sampling of $N_\mathrm{live}$ points that are distributed according to the prior, hence drawn from the prior PDF itself. At each new iteration, a new sampling point is drawn from the prior PDF with a corresponding likelihood value that satisfies the hard constraint $\mathcal{L} > \mathcal{L}^*$, with $\mathcal{L}^*$ the worst likelihood value of the previous iteration. The point associated to the worst likelihood value is then removed from the sample and a new iteration starts. At the end, the prior mass reached corresponds to $X=1$ and the sampling terminates in a region that is located around the maximum (or the maxima) of the likelihood function.

\subsection{The \diamonds\,\,code}
\begin{figure}[t]
   \centering
   \includegraphics[width=6.0cm]{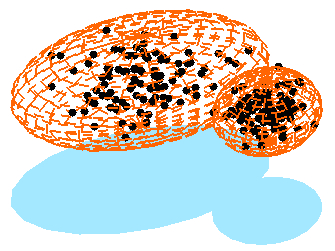}\includegraphics[width=5.0cm]{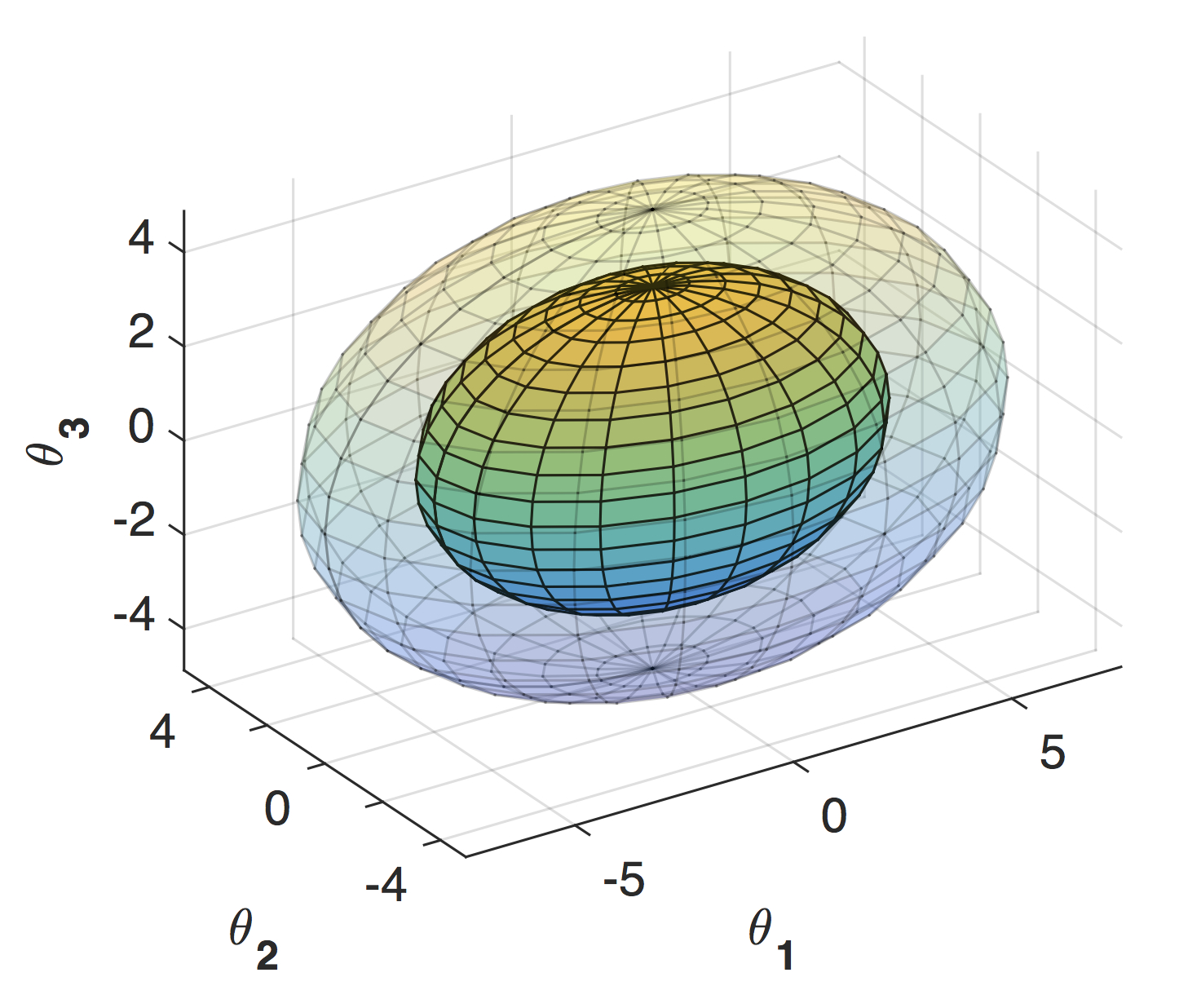}
      \caption{\textit{Left panel:} Three-dimensional ellipsoids containing two different clusters of sampling points in the parameter space. \textit{Right panel:} The enlargement of an ellipsoid used to optimize the sampling efficiency throughout the nesting process.}
    \label{fig:ellipsoid}
\end{figure}

\begin{figure}[t]
   \centering
   \includegraphics[width=10.0cm]{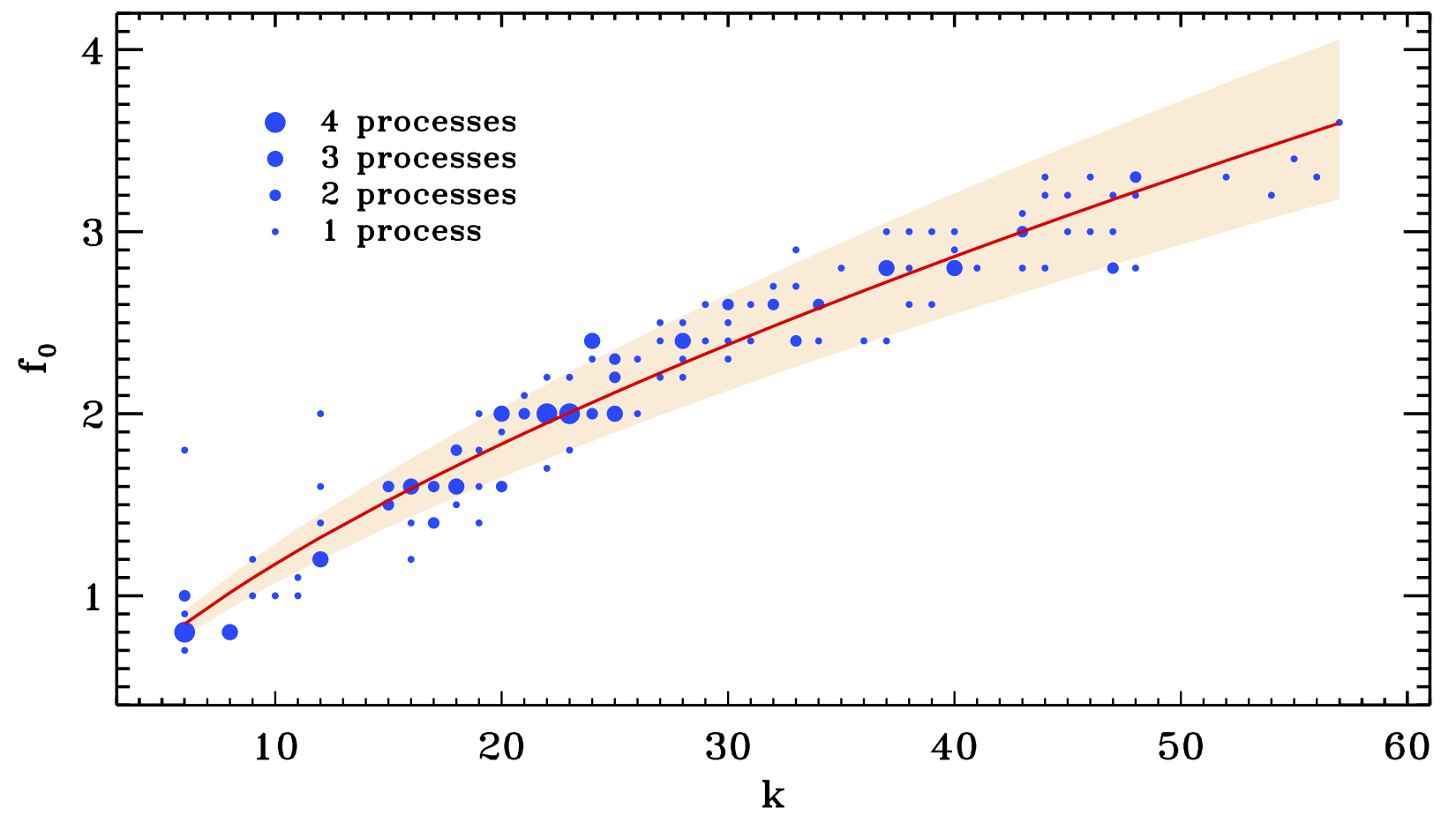}
      \caption{The initial enlargement fraction $f_0$ as a function of the number of dimensions $k$ involved in the inference problem. The 152 independent computations provided by \cite{Corsaro15cat} used 4 clusters each to sample the parameter space. The size of the circles is proportional to the number of processes for which the same $f_0$ was used. The colored band shows the 68.3\,\% confidence region for the power law fit (thick red line).}
    \label{fig:f0}
\end{figure}

The high-DImensional And multi-MOdal NesteD Sampling (\diamonds) code\footnote{DIAMONDS is publicly available at \url{https://fys.kuleuven.be/ster/Software/Diamonds/} or through its public GitHub repository at \url{https://github.com/EnricoCorsaro/DIAMONDS}.} is a {\ttfamily C++11} software for Bayesian parameter estimation and model comparison that uses a version of the NSMC algorithm. A major difficulty in implementing the NSMC algorithm is the drawing from the prior PDF that satisfies the hard constraint in the likelihood value of the drawn point. Following on the developments made for other existing codes that implement NSMC \citep[see, e.g.,][]{Shaw07,Feroz08,Feroz09}, \diamonds\,\,overcomes this problem by adopting a simultaneous ellipsoidal sampling algorithm \citep{Corsaro14}. This means that the posterior PDF is actually sampled by means of multi-dimensional ellipsoids, which decompose the parameter space $\Omega_{\mathcal{M}_k}$ into small hyper-volumes, as shown in the left panel of Fig.~\ref{fig:ellipsoid}. Each ellipsoid can thus be used to easily draw new points from, and it is reduced in its volume as the nested iteration proceeds toward a termination condition. In particular, one crucial parameter to control the behavior of the ellipsoids is the initial enlargement fraction, $f_0$, which is used to enlarge their axes along each direction for as many dimensions as imposed by the number of free parameters. This parameter, whose effect is depicted in the right panel of Fig.~\ref{fig:ellipsoid}, tunes the efficiency of the sampling throughout the nested iterations and therefore requires a careful calibration, which I show in Fig.~\ref{fig:f0} as a function of the number of dimensions, $k$. A calibrated relation, already implemented in \diamonds, reads
\begin{equation}
f_0 = (0.267 \pm 0.014) \, k^{0.643 \pm 0.017}
\end{equation}
and allows for using \diamonds\,\,for a wide range of applications without the need to adjust the parameter $f_0$ every time a new model or a different number of parameters is involved in the analysis.

\diamonds\,\,includes a library of likelihood functions and prior PDFs that can be used for a wide range of applications. As for any inference problem, the code requires an input dataset, a model to be fit to the observations, and the adoption of a given likelihood function and of prior PDFs for each free parameter of the model. The termination condition that allows the code to finalize its computations is based on the remaining Bayesian evidence, as described by \cite{Keeton11} (see also \citealt{Corsaro14} for additional details). Instructions on how to configure the code and a description of its different parts can be found in the online user guide\footnote{A comprehensive user guide to DIAMONDS can be found at \url{https://fys.kuleuven.be/ster/Software/Diamonds/DIAMONDS_UserGuide}.}. In the following examples, \diamonds\,\,is set up in different ways depending on the specific inference problem that is considered.

\section{Fitting the background signal}
\label{sec:4}
The first step in the asteroseismic analysis process is to estimate the background signal in the power spectrum of a star\footnote{The power spectrum is usually converted into a power spectral density, PSD, to allow for direct comparisons independently of the observing length of the data. Its units are expressed in ppm$^2$ $\mu$Hz$^{-1}$}. This is an important phase of the analysis because if not properly performed it can introduce significant systematics in the asteroseismic parameters that characterize individual oscillation modes \citep{Corsaro14}. The first part of the tutorial is therefore focused on the estimation of the background signal in the red giant KIC~12008916, observed by NASA's \textit{Kepler} mission \citep{Borucki10,Koch10} for more than four years. The dataset has been prepared following \cite{Garcia11data,Garcia14}, thus optimized for asteroseismic analysis.

In order to run the tutorial, one needs to have the \diamonds\,\,code already installed in a local machine. This procedure can be accomplished by following the instructions provided in the installation guide section of the code website\footnote{The installation guide of DIAMONDS can be found at \url{https://fys.kuleuven.be/ster/Software/Diamonds/installation-guide}.}. Subsequently it is required to download the code extension for background fitting\footnote{The Background extension of DIAMONDS can be downloaded from \url{https://fys.kuleuven.be/ster/Software/Diamonds/package/AzoresSC16\_background\_extension.tar.gz}. Further information on how to run the tutorial can be found at \url{http://www.iastro.pt/research/conferences/faial2016/files/presentations/TA1.pdf}.}, containing the specific fitting model, priors, and dataset to be used in the tutorial. The extension contains a library of {\ttfamily Python} routines that can be used to plot the results obtained with \diamonds. We note that throughout this tutorial we will adopt an exponential likelihood function, as appropriate for datasets deriving from a Fourier transform of a time series \citep{Duvall86,Corsaro14}.

\begin{figure}[t]
   \centering
   \includegraphics[width=11.0cm]{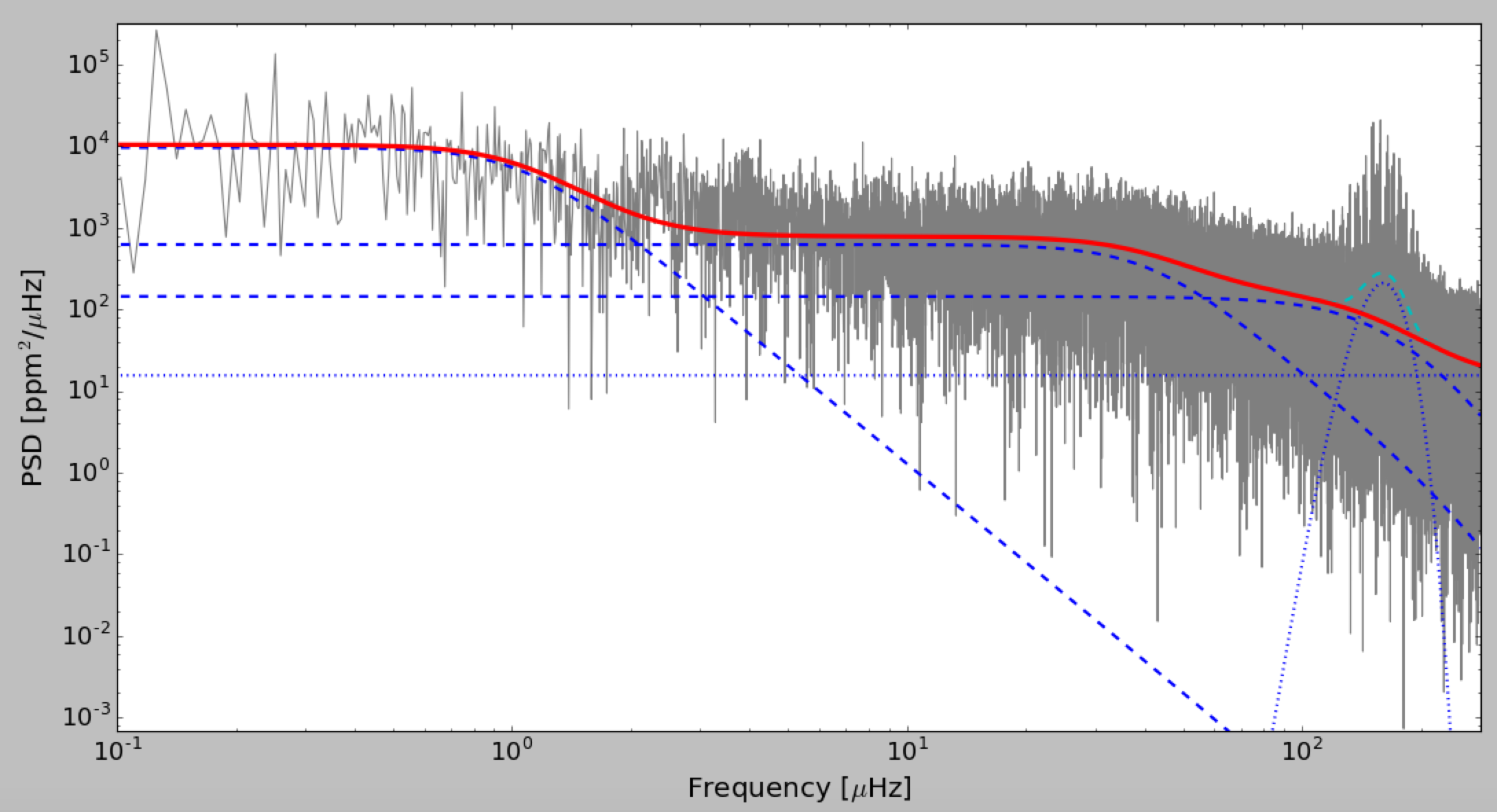}
      \caption{Background fit of the star KIC12008916 by means of \textsc{D\normalsize{iamonds}}. The original PSD is shown in gray. The red thick line represents the background model without the Gaussian envelope. The cyan dotted line accounts for the additional Gaussian component. The individual components of the background model as given by Eq.~(\ref{eq:bkg}) are shown by blue dot-dashed lines.}
    \label{fig:bkg}
\end{figure}

The background model, considered as a function of the cyclic frequency in the PSD of the star, reads
\begin{equation}
P_\mathrm{bkg} \left(\nu \right) = W + R \left( \nu \right) \left[ B\left( \nu \right) + G \left( \nu \right) \right] \, ,
\label{eq:bkg}
\end{equation}
where $W$ is a flat noise level and $R\left(\nu\right)$ the response function that considers the sampling rate of the observations for \textit{Kepler} data,
\begin{equation}
R\left( \nu \right) = \mbox{sinc}^2 \left( \frac{\pi \nu}{2 \nu_\mathrm{Nyq}} \right) \, ,
\label{eq:resp}
\end{equation}
with $\nu_\mathrm{Nyq} = 283.212\,\mu$Hz the Nyquist frequency in the case of long-cadence data \citep{Jenkins10}. We fit three Harvey-like profiles \citep{Harvey85} given by
\begin{equation}
B\left(\nu\right) = \sum^3_{i=1} \frac{\zeta a^2_i / b_i}{1 + \left( \nu / b_i \right)^4} \, ,
\label{eq:bkg_harvey}
\end{equation}
with $a_i$ the amplitude in ppm, $b_i$ the characteristic frequency in $\mu$Hz, and $\zeta = 2\sqrt{2}/\pi$ the normalization constant \citep{Kallinger14}.
The power excess containing the oscillations is described as
\begin{equation}
G\left(\nu\right) = H_\mathrm{osc} \exp \left[ - \frac{ \left( \nu - \nu_\mathrm{max} \right)^2}{2 \sigma_\mathrm{env}^2} \right]
\label{eq:env}
\end{equation}
and is only considered when fitting the background model to the overall PSD of the star. The global model given by Eq.~(\ref{eq:bkg}) therefore accounts for ten free parameters. The resulting fit obtained with \diamonds\,\,is shown in Fig.~\ref{fig:bkg}. 

\begin{svgraybox}
Questions \& Problems:
\begin{itemize}
\item For any of the estimated free parameters, which Bayesian parameter estimator should be preferred among the mode, the median and the mean? And why?
\item What is the value of $\nu_\mathrm{max}$ for this star?
\item Could you guess what the evolutionary stage of this red-giant star is from its $\nu_\mathrm{max}$ value?
\item Using your fitted $\nu_\mathrm{max}$, and assuming $\Delta\nu\!=\!12.9\,\mu$Hz as the large frequency separation \citep{Ulrich86}, $T_\mathrm{eff} = 5100\,$K, and solar reference values $\nu_\mathrm{max,\odot} = 3100\,\mu$Hz, $\Delta\nu_{\odot} = 134.9\,\mu$Hz, and $T_\mathrm{eff,\odot} = 5777\,$K, estimate the mass and radius of the star through scaling relations.
\end{itemize}
\end{svgraybox}

\section{Fitting the oscillation modes}
\label{sec:5}
The second part of the tutorial is related to the fitting of the oscillation modes. For this purpose it is necessary to download and install the extension of \diamonds \,related to the peak-bagging analysis\footnote{The PeakBagging extension of DIAMONDS can be downloaded from \url{https://fys.kuleuven.be/ster/Software/Diamonds/package/AzoresSC16\_peakbagging\_extension.tar.gz}. The extension contains a library of {\ttfamily Python} routines that can be used to plot the results obtained with DIAMONDS. Further informations on how to run the tutorial can be found at \url{http://www.iastro.pt/research/conferences/faial2016/files/presentations/TA1.pdf}.}, similarly to what has been done for the background.

The model that is taken into account is the one presented by \cite{Corsaro15cat} and includes a mixture of resolved and unresolved oscillation mode profiles. For resolved modes, i.e., modes with lifetimes much shorter than the total observing time, the typical profile is a Lorentzian expressed as
\begin{equation}
\mathcal{P}_{\mathrm{res},0} \left( \nu \right) = \frac{A_0^2 / \left( \pi \Gamma_0 \right)}{1 + 4 \left( \frac{\nu - \nu_{0}}{\Gamma_0} \right)^2} \, ,
\label{eq:resolved_profile}
\end{equation}
where $A_0$, $\Gamma_0$, and $\nu_0$ are the amplitude in ppm, the linewidth in $\mu$Hz, and the centroid frequency in $\mu$Hz, respectively, and represent the three free parameters to be estimated during the fitting process. For the unresolved modes, i.e., modes with a lifetime comparable or even longer than the total observing time, we consider the profile
\begin{equation}
\mathcal{P}_{\mathrm{unres},0} \left(\nu \right) = H_0 \, \mbox{sinc}^2 \left[ \frac{\pi \left(\nu - \nu_0 \right)}{\delta \nu_\mathrm{bin}} \right] \, ,
\label{eq:unresolved_profile}
\end{equation}
where $H_0$ and $\nu_0$ are the height in PSD units and the centroid frequency in $\mu$Hz of the oscillation peak, respectively, and must be estimated during the fitting process, while $\delta \nu_\mathrm{bin}$ is fixed as the frequency resolution of the dataset, here corresponding to $0.008$\,$\mu$Hz.

\begin{figure}[t]
  \centering
   \includegraphics[width=11cm]{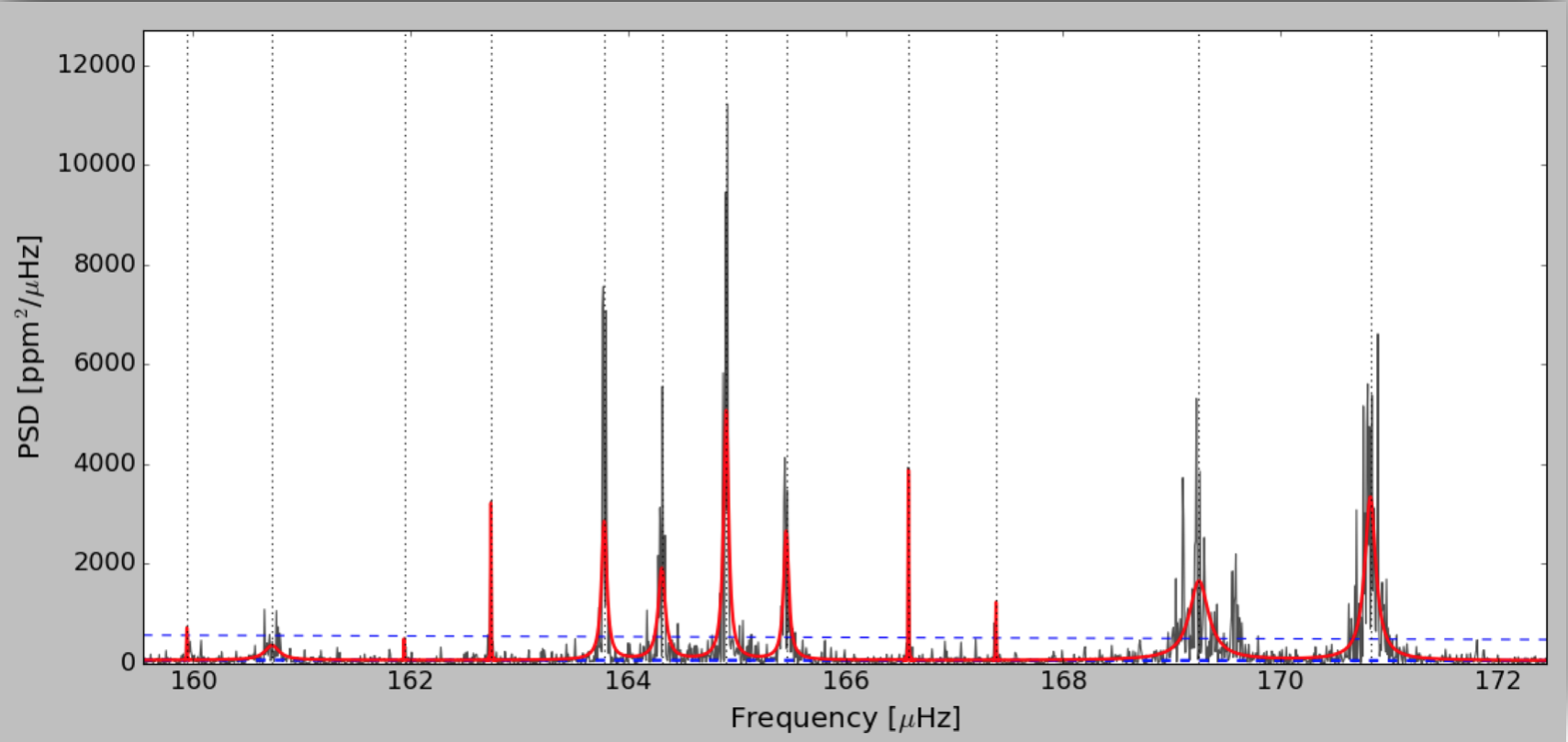}
      \caption{Peak-bagging fit of the star KIC~12008916 by means of \textsc{D\normalsize{iamonds}}. The original PSD is shown in gray. The red thick line represents the estimated peak-bagging model (cf.~Eq.~\ref{eq:general_pb_model}), while the blue dashed lines mark the background signal and a scaled (by a factor of eight) version of it.}
    \label{fig:pkb}
\end{figure}

Following \citet{Corsaro14,Corsaro15cat}, we fix the background parameters corresponding to the white noise, $W = \overline{W}$, and the Harvey-like profiles, $B \left( \nu \right) = \overline{B} \left( \nu \right)$, to the median values estimated in the tutorial in Sect.~\ref{sec:4}. Then, the final peak-bagging model can be represented as
\begin{equation}
P \left( \nu \right) = \overline{W} + R \left(\nu \right) \left[ \overline{B} \left( \nu \right) + P_\mathrm{osc} \left( \nu \right)\right] \, , 
\label{eq:general_pb_model}
\end{equation}
where 
\begin{equation}
P_\mathrm{osc} \left( \nu \right) = \sum^{N_\mathrm{res}}_{i=1} \mathcal{P}_{\mathrm{res},i} \left( \nu \right) + \sum^{N_\mathrm{unres}}_{j=1} \mathcal{P}_{\mathrm{unres},j} \left( \nu \right) \, ,
\label{eq:pb_model}
\end{equation}
with $N_\mathrm{res}$ and $N_\mathrm{unres}$ the number of resolved and unresolved peaks to be fitted, respectively. Clearly, any inference problem that takes into account this peak-bagging model will involve a total number of $3 N_\mathrm{res} + 2 N_\mathrm{unres}$ free parameters. The result of the fit for KIC~12008916 done with \diamonds\,\,is shown in Fig.~\ref{fig:pkb}.

\begin{svgraybox}
Questions \& Problems:
\begin{itemize}
\item In Fig.~\ref{fig:pkb} spot the positions of the radial ($\ell = 0$), quadrupole ($\ell = 2$) and octupole ($\ell = 3$) modes, as follows from the asymptotic relation of the acoustic modes \citep{Tassoul80}.
\item Which oscillation modes are the most p-dominated mixed modes?
\item Compute the spacing (expressed in seconds) between the frequency $\nu_{\ell = 1,m=0} = 165.178\,\mu$Hz and another frequency that has to be computed as the average between the two frequency centroids of the unresolved profiles having the largest frequency (in the range $166$--$168\,\mu$Hz). The frequency centroids of the unresolved profiles must be those from the fitting results obtained with \diamonds.
\item Compare the derived period spacing in the $\Delta P$ - $\Delta \nu$ diagram shown in fig.~8 of \cite{Corsaro12} and determine the evolutionary stage of the star assuming $\Delta\nu\!=\!12.9\,\mu$Hz.
\end{itemize}
\end{svgraybox}

\section{Peak significance test}
\label{sec:6}
As shown by \cite{Corsaro14} and later on applied by \cite{Corsaro15cat} on red-giant stars, by means of the Bayesian evidence it is possible to perform a direct model comparison aimed at assessing the significance of a given oscillation peak. The final part of the tutorial with \diamonds\,\,foresees the computation of the peak significance test for one oscillation mode fitted during the peak-bagging analysis. In order to achieve this result, it is required that the peak-bagging presented in Sect.~\ref{sec:5} is performed with two different models. By selecting a specific oscillation peak that we want to test, then the competing models to be fitted to the PSD of the star have to be defined as follows: (i) the first model, $\mathcal{M}_1$, must contain the entire set of oscillation peaks to be fitted, including the peak that we intend to test; (ii) the second model, $\mathcal{M}_2$, must contain the entire set of peaks to be fitted, except the peak that we intend to test. This implies that the  parameters that configure the prior PDFs of the models $\mathcal{M}_1$ and $\mathcal{M}_2$ should be identical, except for the peak to test. Using the set up of the PeakBagging extension of \diamonds, this can easily be achieved by removing the prior parameters of the corresponding peak when we have to fit model $\mathcal{M}_2$. Among the outputs of \diamonds, there will be the Bayesian evidence\footnote{More details can be found at \url{http://www.iastro.pt/research/conferences/faial2016/files/presentations/TA1.pdf}.}. The best model, or statistically more likely, can be identified by computing the Bayes' factor (see Sect.~\ref{sec:1}) as $\ln \mathcal{B}_{1,2} = \ln \mathcal{E}_1 - \ln \mathcal{E}_2$. If, for example, $\ln \mathcal{B}_{1,2} > 5$, according to Jeffreys' scale of strength for the evidence \citep{Jeffreys61,Trotta08} we then conclude that the peak is significant and that it should be considered as a real oscillation mode.

\begin{svgraybox}
Questions \& Problems:
\begin{itemize}
\item Why are two different models needed to test the significance of an individual peak?
\item How many models are required to test the significance of two peaks?
\item Perform the peak significance test for the $\ell = 3$ mode shown in Fig.~\ref{fig:pkb} by means of \diamonds.
\item Provide the value of the natural logarithm of the Bayes' factor for the aforementioned oscillation mode and assess the strength of the evidence according to Jeffreys' scale.
\end{itemize}
\end{svgraybox}

\begin{acknowledgement}
This work has been funded by the European Community's Seventh Framework Programme (FP7/2007-2013) under grant agreement n$^\circ$312844 (SPACEINN).
\end{acknowledgement}

\bibliographystyle{apj}
\bibliography{biblio}

\begin{thebibliography}{}
\expandafter\ifx\csname natexlab\endcsname\relax\def\natexlab#1{#1}\fi

\bibitem[{Bolstad(2013)}]{Bolstad13}
Bolstad, W. 2013, Introduction to Bayesian Statistics (Wiley)

\bibitem[{{Borucki} {et~al.}(2010){Borucki}, {Koch}, {Basri}, {Batalha},
  {Brown}, {Caldwell}, {Caldwell}, {Christensen-Dalsgaard}, {Cochran},
  {DeVore}, {Dunham}, {Dupree}, {Gautier}, {Geary}, {Gilliland}, {Gould},
  {Howell}, {Jenkins}, {Kondo}, {Latham}, {Marcy}, {Meibom}, {Kjeldsen},
  {Lissauer}, {Monet}, {Morrison}, {Sasselov}, {Tarter}, {Boss}, {Brownlee},
  {Owen}, {Buzasi}, {Charbonneau}, {Doyle}, {Fortney}, {Ford}, {Holman},
  {Seager}, {Steffen}, {Welsh}, {Rowe}, {Anderson}, {Buchhave}, {Ciardi},
  {Walkowicz}, {Sherry}, {Horch}, {Isaacson}, {Everett}, {Fischer}, {Torres},
  {Johnson}, {Endl}, {MacQueen}, {Bryson}, {Dotson}, {Haas}, {Kolodziejczak},
  {Van Cleve}, {Chandrasekaran}, {Twicken}, {Quintana}, {Clarke}, {Allen},
  {Li}, {Wu}, {Tenenbaum}, {Verner}, {Bruhweiler}, {Barnes}, \&
  {Prsa}}]{Borucki10}
{Borucki}, W.~J., {Koch}, D., {Basri}, G., {et~al.} 2010, Science, 327, 977

\bibitem[{{Corsaro} \& {De Ridder}(2014)}]{Corsaro14}
{Corsaro}, E., \& {De Ridder}, J. 2014, A\&A, 571, A71

\bibitem[{{Corsaro} {et~al.}(2015){Corsaro}, {De Ridder}, \&
  {Garc{\'{\i}}a}}]{Corsaro15cat}
{Corsaro}, E., {De Ridder}, J., \& {Garc{\'{\i}}a}, R.~A. 2015, A\&A, 579, A83

\bibitem[{{Corsaro} {et~al.}(2013){Corsaro}, {Fr{\"o}hlich}, {Bonanno},
  {Huber}, {Bedding}, {Benomar}, {De Ridder}, \& {Stello}}]{Corsaro13}
{Corsaro}, E., {Fr{\"o}hlich}, H.-E., {Bonanno}, A., {et~al.} 2013, MNRAS, 430,
  2313

\bibitem[{{Corsaro} {et~al.}(2012){Corsaro}, {Grundahl}, {Leccia}, {Bonanno},
  {Kjeldsen}, \& {Patern{\`o}}}]{Corsaro12}
{Corsaro}, E., {Grundahl}, F., {Leccia}, S., {et~al.} 2012, A\&A, 537, A9

\bibitem[{{Duvall} \& {Harvey}(1986)}]{Duvall86}
{Duvall}, Jr., T.~L., \& {Harvey}, J.~W. 1986, in NATO ASIC Proc. 169:
  Seismology of the Sun and the Distant Stars, ed. D.~O. {Gough}, 105--116

\bibitem[{{Feroz} \& {Hobson}(2008)}]{Feroz08}
{Feroz}, F., \& {Hobson}, M.~P. 2008, MNRAS, 384, 449 (FH08)

\bibitem[{{Feroz} {et~al.}(2009){Feroz}, {Hobson}, \& {Bridges}}]{Feroz09}
{Feroz}, F., {Hobson}, M.~P., \& {Bridges}, M. 2009, MNRAS, 398, 1601 (F09)

\bibitem[{{Garc{\'{\i}}a} {et~al.}(2011){Garc{\'{\i}}a}, {Hekker}, {Stello},
  {Guti{\'e}rrez-Soto}, {Handberg}, {Huber}, {Karoff}, {Uytterhoeven},
  {Appourchaux}, {Chaplin}, {Elsworth}, {Mathur}, {Ballot},
  {Christensen-Dalsgaard}, {Gilliland}, {Houdek}, {Jenkins}, {Kjeldsen},
  {McCauliff}, {Metcalfe}, {Middour}, {Molenda-Zakowicz}, {Monteiro}, {Smith},
  \& {Thompson}}]{Garcia11data}
{Garc{\'{\i}}a}, R.~A., {Hekker}, S., {Stello}, D., {et~al.} 2011, MNRAS, 414,
  L6

\bibitem[{{Garc{\'{\i}}a} {et~al.}(2014){Garc{\'{\i}}a}, {Mathur}, {Pires},
  {R{\'e}gulo}, {Bellamy}, {Pall{\'e}}, {Ballot}, {Barcel{\'o} Forteza},
  {Beck}, {Bedding}, {Ceillier}, {Roca Cort{\'e}s}, {Salabert}, \&
  {Stello}}]{Garcia14}
{Garc{\'{\i}}a}, R.~A., {Mathur}, S., {Pires}, S., {et~al.} 2014, A\&A, 568,
  A10

\bibitem[{{Harvey}(1985)}]{Harvey85}
{Harvey}, J. 1985, in ESA Special Publication, Vol. 235, Future Missions in
  Solar, Heliospheric \& Space Plasma Physics, ed. E.~{Rolfe} \& B.~{Battrick},
  199--208

\bibitem[{Jeffreys(1961)}]{Jeffreys61}
Jeffreys, H. 1961, Theory of Probability (3rd Ed. OUP Oxford)

\bibitem[{{Jenkins} {et~al.}(2010){Jenkins}, {Caldwell}, {Chandrasekaran},
  {Twicken}, {Bryson}, {Quintana}, {Clarke}, {Li}, {Allen}, {Tenenbaum}, {Wu},
  {Klaus}, {Van Cleve}, {Dotson}, {Haas}, {Gilliland}, {Koch}, \&
  {Borucki}}]{Jenkins10}
{Jenkins}, J.~M., {Caldwell}, D.~A., {Chandrasekaran}, H., {et~al.} 2010, ApJ,
  713, L120

\bibitem[{{Kallinger} {et~al.}(2014){Kallinger}, {De Ridder}, {Hekker},
  {Mathur}, {Mosser}, {Gruberbauer}, {Garc{\'{\i}}a}, {Karoff}, \&
  {Ballot}}]{Kallinger14}
{Kallinger}, T., {De Ridder}, J., {Hekker}, S., {et~al.} 2014, A\&A, 570, A41

\bibitem[{{Keeton}(2011)}]{Keeton11}
{Keeton}, C.~R. 2011, MNRAS, 414, 1418 (K11)

\bibitem[{{Koch} {et~al.}(2010){Koch}, {Borucki}, {Basri}, {Batalha}, {Brown},
  {Caldwell}, {Christensen-Dalsgaard}, {Cochran}, {DeVore}, {Dunham},
  {Gautier}, {Geary}, {Gilliland}, {Gould}, {Jenkins}, {Kondo}, {Latham},
  {Lissauer}, {Marcy}, {Monet}, {Sasselov}, {Boss}, {Brownlee}, {Caldwell},
  {Dupree}, {Howell}, {Kjeldsen}, {Meibom}, {Morrison}, {Owen}, {Reitsema},
  {Tarter}, {Bryson}, {Dotson}, {Gazis}, {Haas}, {Kolodziejczak}, {Rowe}, {Van
  Cleve}, {Allen}, {Chandrasekaran}, {Clarke}, {Li}, {Quintana}, {Tenenbaum},
  {Twicken}, \& {Wu}}]{Koch10}
{Koch}, D.~G., {Borucki}, W.~J., {Basri}, G., {et~al.} 2010, ApJ, 713, L79

\bibitem[{{Shaw} {et~al.}(2007){Shaw}, {Bridges}, \& {Hobson}}]{Shaw07}
{Shaw}, J.~R., {Bridges}, M., \& {Hobson}, M.~P. 2007, MNRAS, 378, 1365 (S07)

\bibitem[{Sivia \& Skilling(2006)}]{Sivia06}
Sivia, D., \& Skilling, J. 2006, Data Analysis: A Bayesian Tutorial, Oxford
  science publications (OUP Oxford)

\bibitem[{Skilling(2004)}]{Skilling04}
Skilling, J. 2004, AIP Conference Proceedings, 735, 395 (SK04)

\bibitem[{{Tassoul}(1980)}]{Tassoul80}
{Tassoul}, M. 1980, ApJS, 43, 469

\bibitem[{{Trotta}(2008)}]{Trotta08}
{Trotta}, R. 2008, Contemporary Physics, 49, 71

\bibitem[{{Ulrich}(1986)}]{Ulrich86}
{Ulrich}, R.~K. 1986, ApJ, 306, L37

\end{thebibliography}

\end{document}